\documentclass[runningheads]{svmult}

\usepackage{makeidx}   % allows index generation
\usepackage{graphicx}  % standard LaTeX graphics tool
\usepackage{subeqnar}  % subnumbers individual equations
\usepackage{multicol}  % used for the two-column index
\usepackage{physprbb}  % modified textarea for proceedings,
\makeindex             % used for the subject index
\begin{document}
\title*{Detecting filaments at z$\approx$3}
\titlerunning{Detecting filaments at z$\approx$3}
\author{Johan P.U. Fynbo\inst{1}
\and Palle M\o ller\inst{1}
\and Bjarne Thomsen \inst{2}
}

\authorrunning{Fynbo, M\o ller and Thomsen}

\institute{ESO, Garching, Germany \\
\and IFA, \AA rhus, Denmark
}

\maketitle              % typesets the title of the contribution

\begin{abstract}
We present the detection of a filament of Ly$\alpha$ emitting galaxies 
in front of the quasar Q1205-30 at z=3.04 based on deep narrow band
imaging and follow-up spectroscopy obtained at the ESO NTT and VLT. We 
argue
that Ly$\alpha$ selection of high redshift galaxies with relatively
modest amounts of observing time allows the detection and redshift 
measurement of galaxies with sufficiently high space 
densities that we can start to map out the large scale structure at
z$\approx$2-3 directly. Even more
interesting is it that a 3D map of the filaments will provide a
new cosmological test for the value of the cosmological constant,
$\Omega_{\Lambda}$.
\end{abstract}

\section{Introduction}
\begin{figure}[ht]
\begin{center}
\includegraphics[width=0.9\textwidth,clip=]{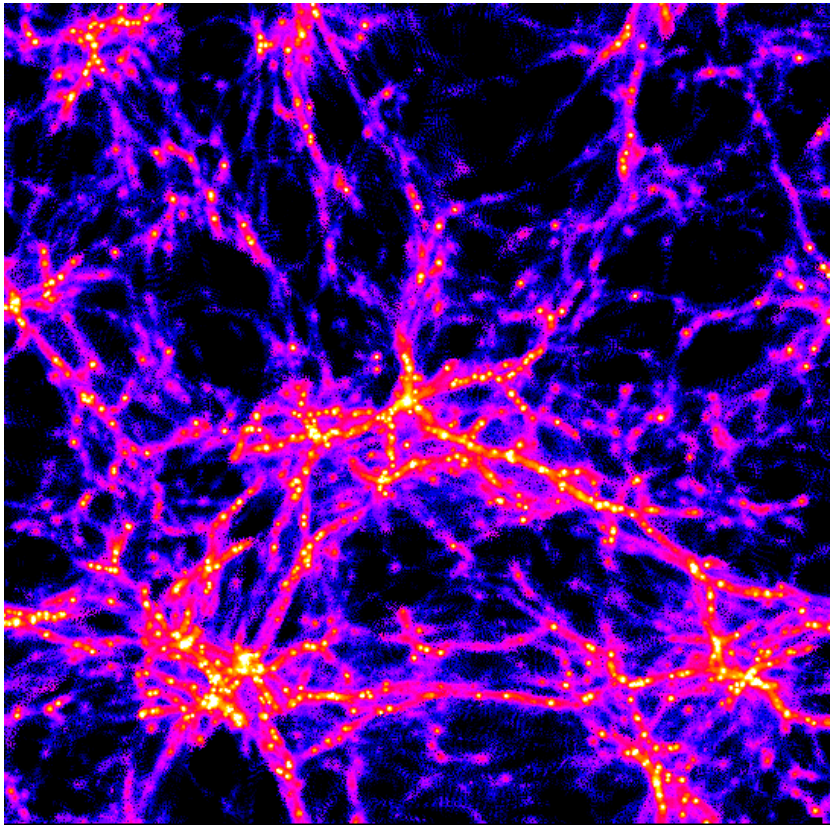}
\end{center}
\caption[]{A hydro simulation of a region of comoving size 
12.5/h$\times$12.5/h$\times$1.6/h Mpc$^3$ at z=3 (courtesy of Tom
Theuns, IoA, Cambridge). The image
shows the density distribution. The white regions, corresponding 
to overdensities above 100, are distributed in filamentary 
structures and in the filament intersections.
}
\label{filament}
\end{figure}

For the past few decades computer simulations have been ahead of the
observations when it comes to describing the first structures to form
at high redshifts. The present consensus of the model builders is
that the gas arranges
itself in long string--like structures commonly referred to as filaments
(see Fig.~\ref{filament}).
Density variations along the filaments will lead to formation of
lumps of cold, self--shielding HI regions and those regions are
identified, in the simulations, as regions of starformation. Because
of the high column density of neutral Hydrogen a sightline through
such a cloud
intersects, they are also identified as strong absorbers known as
Damped Ly$\alpha$ Absorbers (DLAs). By poking random sightlines through
a virtual universe one may simulate observations, and a given
model universe will hence predict a
specific correlation between DLA systems and the galaxies hosting the
DLAs (e.g. Katz et al. 1996). Comparison to real observations of DLA 
galaxies (M\o ller \& Warren 1998) has shown that there is 
very good agreement between observations and simulations. This agreement is
encouraging, but it would be of great interest if one could observationally 
map out the actual filaments. Until now this his been done only at low 
redshifts (e.g. De Lapparent et al. 1991; Bharadwaj  et al. 2000), but never 
at z$>$0.1. Knowing the distribution of scalesizes 
of filaments at different redshifts will help constrain
the allowable parameter space of the simulations. 

Unfortunately such a map cannot be constructed directly via absorption 
studies, because there is currently not a sufficiently tight mesh of 
background z$>$3 quasars available (e.g. Pichon et al. 2001). The best 
way to proceed is hence to attempt to find enough
centres of starformation to be able to map out filaments by their own
light. In order to identify objects to map out filaments one might
at first guess that a search for Lyman Break Galaxies
(LBGs, Steidel \& Hamilton 1992) would be the best procedure.
Unfortunately only the very brightest galaxies can be found and have their 
redshifts measured precisely enough with
this technique, and such sparse sampling of the filamentary structure
does not allow the structures to be seen. However, it has been shown
that both DLA galaxies and galaxies selected for their Ly$\alpha$
emission, are sampling the high redshift
galaxy population much further down the Luminosity function than do
the LBGs, and one will therefore expect a better sampling of the
high redshift structure if one uses DLA galaxies and Ly$\alpha$
galaxies (Fynbo, M\o ller \& Warren 1999; Haehnelt et al. 2000). 
This has recently 
been independently confirmed, as deep narrow band Ly$\alpha$ imaging in 
a known overdensity of LBGs revealed about a factor
of 10 more candidate Ly$\alpha$ galaxies than LBGs (Steidel et al.
2000).

\section{Observations}
\begin{figure}[ht]
\begin{center}
\includegraphics[width=0.7\textwidth]{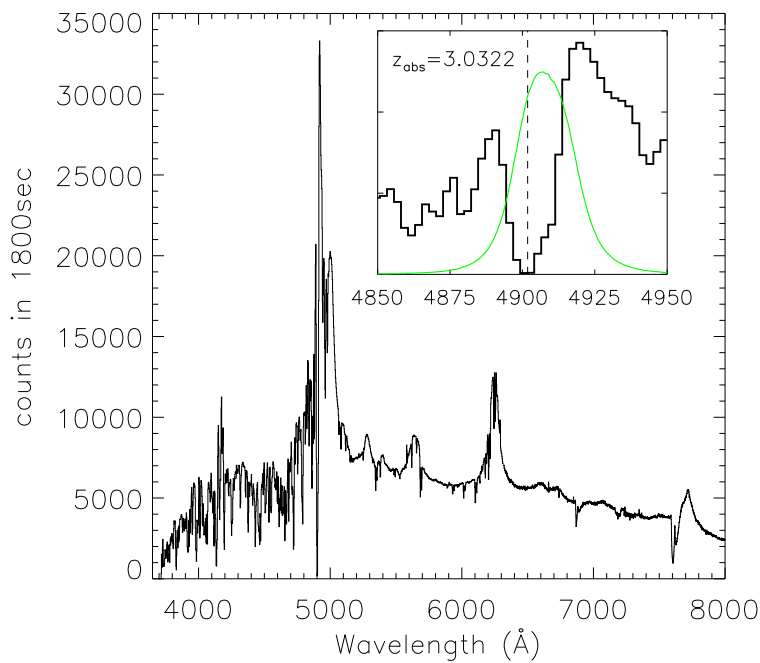}
\end{center}
\vskip -2cm
\caption[]{The spectrum of Q1205-30 obtained at the VLT in March 2000. 
The insert in the uper right hand corner shows the region of the spectrum 
around a strong Ly$\alpha$ absorption line z=3.0322. The narrow
filter (transmission curve overplotted) was tuned to Ly$\alpha$ at
this redshift. The redshift of the background QSO, as measured from the 
low ionization OI line, is z=3.0473$\pm$0.0012.}
\label{QSO}
\end{figure}

In February through March 1998 we obtained deep narrow band imaging
in a 21\AA \ wide filter tuned to Ly$\alpha$ at z=3.04. The filter was
tuned to the wavelength of a strong Ly$\alpha$ absorption line
in the spectrum of the QSO (Fig.~\ref{QSO}). 
The data were collected as service observing program at the 3.5-m ESO 
New Technology Telescope on La Silla, Chile. In total almost 18 hours 
of narrow band imaging was secured reaching a 5$\sigma$ flux limit of 
1.1$\times$10$^{-17}$ erg s$^{-1}$ cm$^{-2}$ (Fynbo, Thomsen \& M\o ller 
2000). 
We detected six good ($>$5$\sigma$) and two marginal ($\sim$4$\sigma$) 
candidate Ly$\alpha$ emitters in 
the field of the QSO as well as extended Ly$\alpha$ emission close to 
the QSO line of sight. In March 2000 we obtained Multi-Object follow-up 
spectroscopy at the ESO Very Large Telescope using FORS1 on the UT1 unit 
telescope. We also obtained deep imaging in the B and I bands (reaching
B(AB)=26.7 and I(AB)=25.9 at 5$\sigma$).

\section{Results}

\begin{figure}[ht]
\begin{center}
\includegraphics[width=1.0\textwidth,clip=]{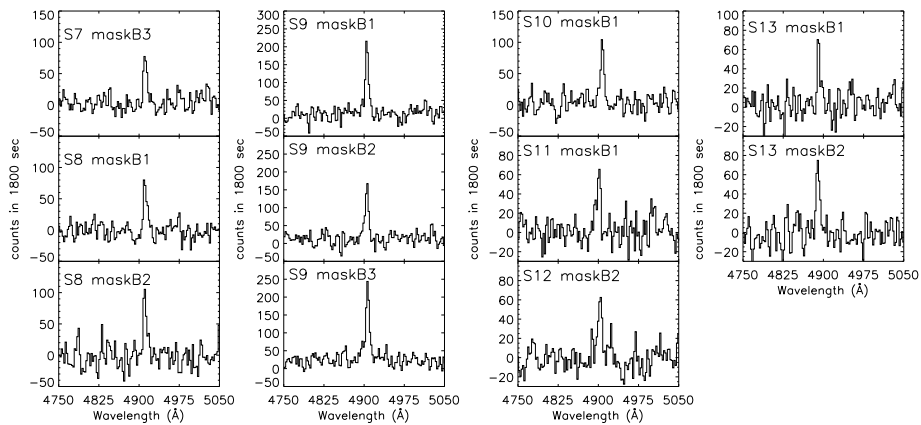}
\end{center}
\caption[]{
The spectral regions around the Ly$\alpha$ emission lines for the
7 confirmed Ly$\alpha$ emitters (named S7--S13). For S8, S9, and S13 
we show the spectra from several masks. As seen, the presence of 
emission lines is confirmed for all 7 candidates. By obtaining
spectra in a red grism covering the spectral region from 5500\AA \
to 7500\AA \ we exclude the possibility of low redshift interlopers
(e.g. OII emiters at z$\approx$0.3, see Fynbo, M\o ller \& Thomsen 2001
for details).
}
\label{spec}
\end{figure}

The VLT spectroscopy confirmed (by detecting the Ly$\alpha$ line and
at the same time excluding the possibility of low redshift interlopers) 
all six good candidates and one of the 
two marginal candidates as Ly$\alpha$ emitters at z=3.04 (Fynbo, 
M\o ller \& Thomsen 2001). The spectral regions around Ly$\alpha$ for 
all confirmed candidates 
are shown in Fig.~\ref{spec}. The spectroscopy of the extended emission close 
to the QSO line-of-sigt will be presented in a separate paper (Weidinger et 
al. in preparation).

In Table~\ref{redtab} we present the redshifts and celestial
positions for all confirmed Ly$\alpha$ emitters and for the absorber
(from M\o ller \& Fynbo 2001). 

%=====================Begin Table 1==============================
\begin{table}
\begin{center}
\caption{Redshifts and positions of seven Ly$\alpha$ emitters and a
Ly$\alpha$ absorber in the field of Q1205--30. The positions are
given relative to the quasar coordinates:
12:08:12.7, -30:31:06.10 (J2000.0). The uncertainty on the redshifts
is 0.0012 (1$\sigma$).}
\begin{tabular}{@{}lrrl}
\hline
Object & $\Delta$RA (arcsec) & $\Delta$decl. (arcsec) & redshift \\
\hline
S7  & -143.3$\pm$0.6 &  41.9$\pm$0.2 & 3.0402  \\
S8  & -141.5$\pm$0.6 &  59.7$\pm$0.2 & 3.0398  \\
S9  & -124.6$\pm$0.5 &  63.4$\pm$0.2 & 3.0350  \\
S10 & -119.9$\pm$0.5 &  59.8$\pm$0.2 & 3.0353  \\
S11 &  -77.8$\pm$0.3 &   0.9$\pm$0.1 & 3.0312  \\
S12 &  -43.9$\pm$0.2 &  54.4$\pm$0.2 & 3.0333  \\
S13 &   68.3$\pm$0.3 & -52.1$\pm$0.2 & 3.0228  \\
abs &    0.0         &   0.0         & 3.0322  \\
\hline
\label{redtab}
\end{tabular}
\vspace{-1cm}
\end{center}
\end{table}

\subsection{Filamentary structure}
\begin{figure}[ht]
\begin{center}
\includegraphics[width=1.0\textwidth, clip=]{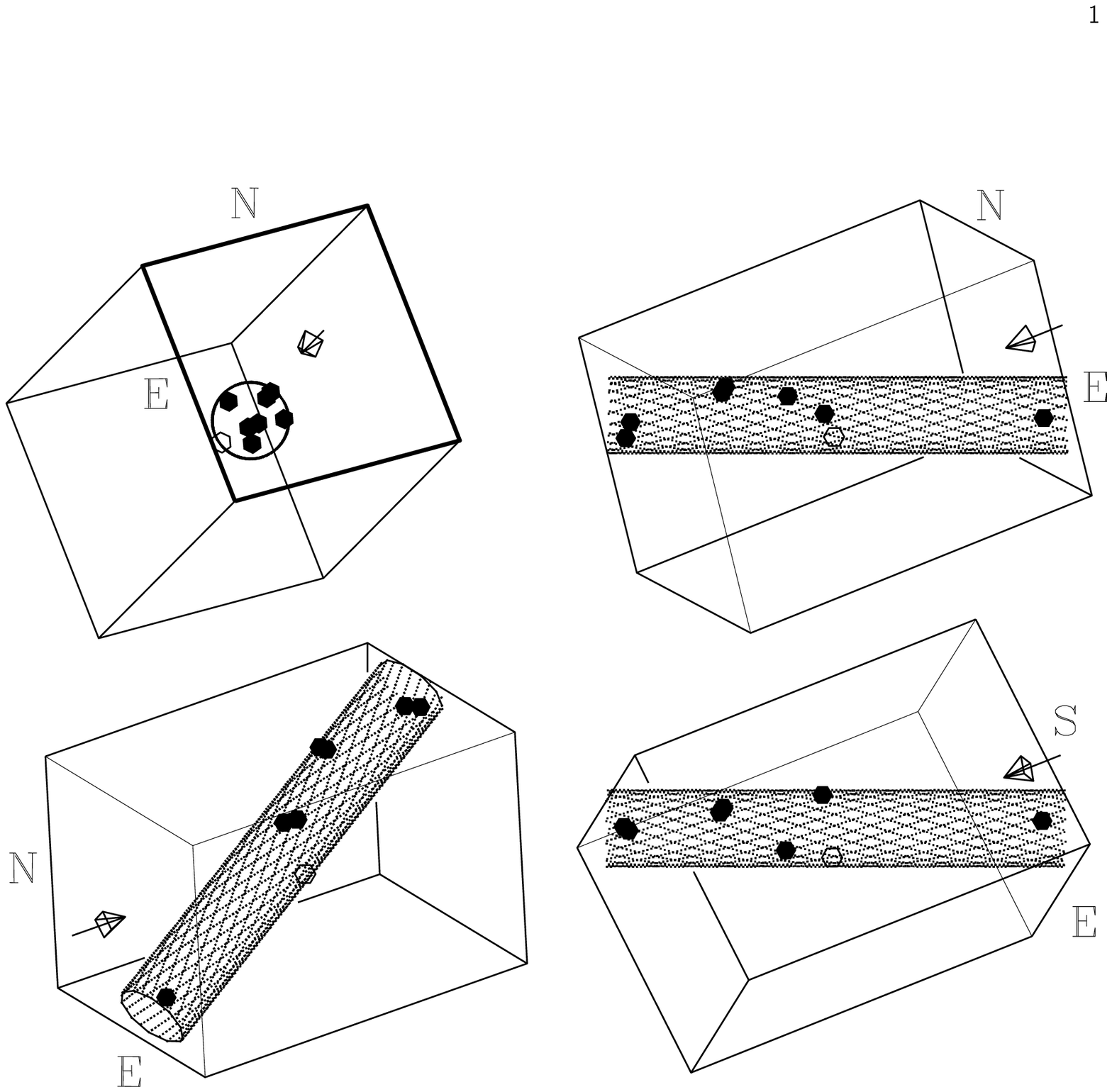}
\end{center}
\vskip -3cm
\caption[]{
3D distribution of the eight objects seen from 4 different
viewing--angles.
In each of the figures the 3D arrow points in our viewing
direction on the sky, and the spiral pattern maps out a cylinder with
radius 400 kpc (see Table~\ref{redtab}). The box marks the volume of 
space observed with our narrow--band Ly$\alpha$ filter.
{\bf Top left:}
Here we have rotated the view to look along the filament. The thick
lines mark the front "entrance window" of the box (corresponding to
our CCD image).
{\bf Top right:}
The box is here rotated 90 degrees to the right, hence viewing the
filament from the left side compared to the end--on view.
{\bf Bottom right:}
Same as top right but rotated 90 degrees around the filament to give a
view of the filament as seen from ``above'' the view in top left.
{\bf Bottom left:}
View from a random angle to give an impression of the 3D structure.
}
\label{filafig}
\end{figure}

In Fig.~\ref{filafig} we show the objects plotted in the box defined 
by the Field of View of the Camera and
the redshift depth of the filter for Ly$\alpha$ at z=3. As seen 
the 7 Ly$\alpha$ emitters (marked with filled symbols) and the absorber 
(the open symbol) all align in this diagram. If we assume that the
redshifts are all solely due to Hubble flow then this implies a real 
alignment in 3D space e.g. a filamentary spatial distribution of the 
objects. However, the measured redshift may not be due to Hubble flow
alone for mainly two effects : {\it i)} outflows, and {\it ii)} peculiar 
velocities. 
We now briefly discuss the importance of each of these effects. 
{\it Outflows:} In the nearby starburst galaxy NGC1705 the outflow 
velocity is estimated to be around 80 km s$^{-1}$ (Heckman et al. 2001 
and references therein). Outflows of this strength will cause a
shift in the redshift measurement which is of the same order as 
the combined uncertainty from the wavelength calibration and line 
centroid measurement (corresponding to 90 km s$^{-1}$ at z=3). 
Outflows will either produce a systematic blueshift of the emission line
redshift if the galaxies are opaque (so that we only see the gas moving 
towards us) or a broadening with no velocity shift of the lines (if the 
galaxies are transparent and we also see the gas moving away from us).
The fact that the absorber, for which the redshift is detemined from the 
Ly$\alpha$ absorption line, also follows the alignment is an argument 
against a significant blueshift due to outflows. 
{\it Peculiar velocities:} In the local universe
(v$<$4000 km s$^{-1}$) the $\sigma$(v$_{peculiar}$) of peculiar 
velocities is of the order 200 km s$^{-1}$ (e.g. Branchini et al. 2001). 
We do not expect this number to be larger at z=3. Furthermore, any 
peculiar velocities will tend to smear out any underlying filamentary 
structure, so the fact that we see alignment is an argument against 
large peculiar velocities.
We therefore conclude that the most likely interpretation of
Fig.~\ref{filafig} is that we see a redshift z=3 filament.

\subsection{Properties of the filament}
We can only determine a lower limit to the length of the filament 
as it seems to extend beyond the volume mapped by our instrumental
setup (CCD and filter). Assuming a Hubble constant of 65 km s$^{-1}$
Mpc$^{-1}$, $\Omega_m=0.3$, and $\Omega_{\Lambda}=0.7$ we find a 
coming length (defined as the distance between the two outhermost
objects) of 4800 proper kpc. The radius of the minimum cylinder
containing all objects is 400 proper kpc. Due to the effect of
peculiar and outflow velocities this radius should be considered
an upper limit.

The derived properties of filaments are strongly dependend on the
assumed cosmology.
In particular, since filaments are anchored in the Hubble flow,
the observed angular distribution of a sample of filaments will
be a function of the assumed cosmology. Therefore, it is in 
principle possible to use a sample of filaments to obtain an
independend constrain on the value of the cosmological constant
at z$\approx$3 (Weidinger et al. 2001).

\section{Summary and outlook}
In order to start mapping out the large scale filamentary structure 
suggested by numerical simulations directly at high redshift we need
cosmic sources that are very numerous rather than rare, very
bright light houses such as QSOs or Gamma-Ray Bursters. We have here
demonstrated that by reaching flux limits below 1$\times$10$^{-17}$
erg s$^{-1}$ cm$^{-2}$ the density of Ly$\alpha$ emitting galaxies 
is sufficiently high at z=3 to allow a direct mapping of filamentary 
structure.

The next logical step is to try to map out larger regions of the
z$\approx$2--3 universe with Ly$\alpha$ emitters. Therefore we
(M\o ller, Fynbo, Thomsen, Egholm, Weidinger, Haehnelt, Theuns)
have initiated a large area survey for Ly$\alpha$ emitters at z=2 
with the 2.56-m Nordic Optical Telescope on La Palma.
Furthermore, in a pilot project conducted at the ESO VLT we (here 
Fynbo, Ledoux, Burud, Leibundgut, M\o ller and Thomsen) have obtained 
narrow band observations of two fields around QSO absorbers at 
z$\approx$3. In the field of the z=2.85 absorber towards Q2138-4427, 
for which our imaging observations are complete, we reach a detection 
limit of about 7$\times$10$^{-18}$ erg s$^{-1}$ cm$^{-2}$ and detect 
34 candidate Ly$\alpha$ emitters in a 45 arcmin$^2$ field over a redshift 
range of $\Delta$z=0.05. This shows that the density of z=3.04 Ly$\alpha$ 
emitters in the Q1205-30 field is not unusually high. Follow-up spectroscopic
observations of the Q2138-4427 field has not yet been obtained. In the
future we hope to map out a large volume with several hundred z=3 
Ly$\alpha$ emitters with the VLT.

\section*{Acknowledgments}
This paper is based on observations collected at the European Southern
Observatory, La Silla and Paranal, Chile (ESO project No. 60.B-0843
and 64.O-0187).

\end{document}